\documentclass[reprint,aps,prb]{revtex4-1}
\usepackage{color,graphicx,calc,url}
\usepackage{dcolumn}
\usepackage{bm}
\usepackage{amssymb}
\usepackage{amsmath}
\usepackage{wasysym}
\usepackage{mathrsfs} 
\usepackage{gensymb}
\usepackage{amssymb}
\usepackage{wrapfig}
\usepackage{floatrow}
\usepackage[ppl]{mathcomp}
\usepackage{color}
\usepackage{MnSymbol}
\definecolor{blue}{rgb}{0,0,1}
\definecolor{red}{rgb}{1,0,0}
\definecolor{darkred}{rgb}{0.5,0,0}

\begin{document}
	\title{Non-local spin Seebeck effect in the bulk easy-plane antiferromagnet NiO}
	
	
	\author{Geert R. Hoogeboom and Bart J. van Wees\email{Electronic mail: g.r.hoogeboom@rug.nl}}
	\affiliation{Physics of Nanodevices, Zernike Institute for Advanced Materials, University of Groningen, Nijenborgh 4, 9747 AG Groningen, The Netherlands.\\
}
	\
	\date{\today}

\begin{abstract}
We report the observation of magnon spin currents generated by the Spin Seebeck effect (SSE) in a bulk single crystal of the easy-plane antiferromagnet NiO. A magnetic field induces a non-degeneracy and thereby an imbalance in the population of magnon modes with opposite spin. A temperature gradient then gives rise to a non-zero magnon spin current. This SSE is measured both in a local and a non-local geometry at 5$\,$K in bulk NiO
. 
The magnetic field dependence of the obtained signal 
is modelled by magnetic field splitting of the low energy magnon modes, affecting the spin Seebeck coefficient. The relevant magnon modes at this temperature are linked to cubic anisotropy and magnetic dipole-dipole interactions. The non-local signal deviates from the expected quadratic Joule heating by saturating at a current from around 75$\,\mu A$ in the injector. 
The magnon chemical potential does not decay exponentially with distance and inhomogeneities may be the result of local magnon accumulations.
%
%


\end{abstract}
\maketitle

Magnon spintronics is a field where spin currents are carried by magnons that exist in tunable magnets for information processing \cite{Chumak2015}. Generation of spin currents in magnets is feasible by using the spin Hall effect (SHE) in a metal injector strip creating the magnons which travel trough the magnetic material to be subsequently detected at a second detecting strip \cite{Cornelissen2015}. Antiferromagnets (AFMs) do not posses stray fields and can therefore be exploited over a wide range of parameters such as external magnetic field and device size. Recently, this non-local technique has effectively been employed for the uniaxial AFMs $\alpha$-Fe$_2$O$_3$ \cite{Lebrun2018a}, MnPS$_3$ \cite{Xing2019} and Cr$_2$O$_3$ \cite{Yuan2018}. Despite their potential for spin transport by both magnons and spin superfluidity \cite{Qaiumzadeh2017}, this geometry has not been employed for easy-plane antiferromagnets like NiO. 

Magnons are quasi particles carrying spin angular momentum which enables the transfer of spins in (insulating) magnets as waves of spin rotations of the magnetic moments. An easy-axis antiferromagnet has left-handed ($\alpha$) and right-handed ($\beta$) magnon modes which energies are equal but which spins are opposite. %
%
%
Magnon interconversion is expected to equal the respective magnon chemical potentials $\mu_m^\alpha=\mu_m^\beta=\mu_m$; the deviation from the equilibrium magnon population. Magnon injection then creates a finite $\mu_m$  which drives the transport of magnon spins, following the regular discussion for magnon transport \cite{Cornelissen2016}. This description is equivalent to Refs. \cite{Shen2019, Flebus2019} where $\mu_\alpha$ and $\mu_\beta$ are regarded as equal but opposite as resulting in opposite spin currents. 
Magnon spins can be injected at the interface with a paramagnetic heavy metal (HM) using the SHE or in the bulk magnet.%

In the first method, $\alpha$ ($\beta$) magnons are created (annihilated) if the accumulated spin direction at the interface is parallel to an $\alpha$ magnon spin resulting in an increase (decrease) is the magnon chemical potential $\mu_m$. 
%
%
%
%
%
In the latter method, heating by the injector sets up a thermal magnon current of both modes, which diffuses from hot to cold region; the spin Seebeck effect. 
Since there is no inherent population imbalance between the modes, there is no net spin current since they carry equal, but opposite spin currents. A magnetic field lifts the degeneracy of these magnon modes, creating an imbalance in their population and thereby net magnon spin currents can be created \cite{Rezende2016, Wu2016}.

The different modes are coupled via inelastic magnon-magnon scattering, allowing for energy interchange between different modes and magnon relaxation. This results in a slight suppression of the spin Seebeck coefficient, but will largely leave the transport and its magnon conductivities of both magnon modes intact \cite{Troncoso2020}. 
The magnon depletion is expected to decay exponentially with distance in bulk magnets
\cite{Shan2016,Troncoso2020}. Accumulation of magnons at interfaces can be observed as a sign change in $\mu_m$ \cite{Shan2016}.

In this letter, the non-local observation of the spin Seebeck effect in an easy-plane AFM shows that there is a imbalance between population of the magnon modes which results in a net spin current driven by the temperature gradient. This is achieved at 5$\,$K in bulk NiO containing multiple domains and there is no need of aligning the magnetic moments by the exchange interaction with a FM layer. The magnon chemical potential shows some local variation and shows an increase in noise by increasing the distance from the injector. The SSE amplitude as a function of the magnetic field strength is modelled by magnon mode splitting, creating an imbalance in the magnon population of the modes. Transport of electrically injected magnons have not been observed in the detector.

\begin{figure}[t]
\includegraphics[width=8.5cm]{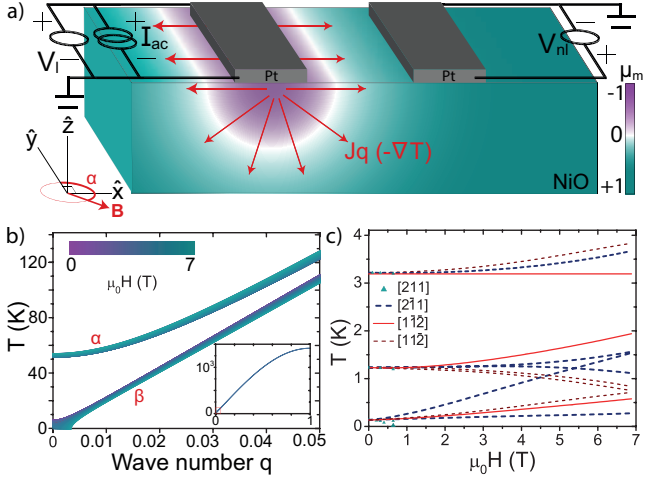}
	\caption{(a) Device structure and probes on the NiO bulk crystal with the spin injector (left Pt strip) and detector (right Pt strip) at a distances d apart, ranging from 250$\,$nm to 7$\,\mu$m. An in-plane magnetic field $\mathbf{B}$ is applied with an angle $\alpha$. A 100$\,\mu$A current is send throught the injector leading to Joule heating and a radial heat gradient J$_q$ indicated by the red arrows. A surplus of magnons at the hot side flow to the cold side, leaving behind a negative $\mu_m$. At interfaces, magnons accumulate and can contribute to $\mu_m$ as observed in thin films of YiG on GGG which distribution is reproduced here \cite{Shan2016}. $\mu_m$ is normalized in the scale. A finite $\mu_m$ results in spin transport between the Pt and the NiO via the spin mixing conductance. The inverse spin Hall effect consequently causes a voltage V$_l$ locally and V$_{nl}$ non-locally. 
(b) The magnon dispersion (Eq. \ref{eq:dispersion_Rezende}) after \cite{Rezende2019} of two modes as a function of a magnetic field along the easy plane considering exchange and Zeeman interaction. The inset shows the full range of the wave number q. (c) Magnon energy at the q=0 point after \cite{Milano2010} as a function of a magnetic field along [110] when taking magnetic dipole-dipole interactions and cubic anisotropy into account. This breaks the symmetry and splits the magnon energies with [2$\overline{\rm 1}$1] spin directions.}
	\label{fig:SSE}
\end{figure}

\begin{figure*}
\floatbox[{\capbeside\thisfloatsetup{capbesideposition={right,top},capbesidewidth=6.0cm}}]{figure}[\FBwidth]
{\caption{(a) Spin Hall magnetoresistance as a function of the in-plane rotation angle $\alpha$, performend at 6.3$\,$T and 5$\,$K. R$_\alpha$ is the angle-dependent resistance of the Pt strip and R$_0$ is the base resistance of 5.17$\,$k$\ohm$. (b) Amplitude of the SMR signals as a function of the magnetic field strength, showing a similar curve as reported in \cite{Hoogeboom2017} for both devices. (c) The locally observed signal changes from the spin Seebeck effect as a function of the magnetic field angle at 6.3$\,$T. (d) The amplitude of the signal increases monotoneously with field with little offset, modelled by the three methods described in the main text. The error bars in (b) and (d) represent the fit uncentainty.}
	\label{fig:locally}}
{\includegraphics[width=9.4cm]{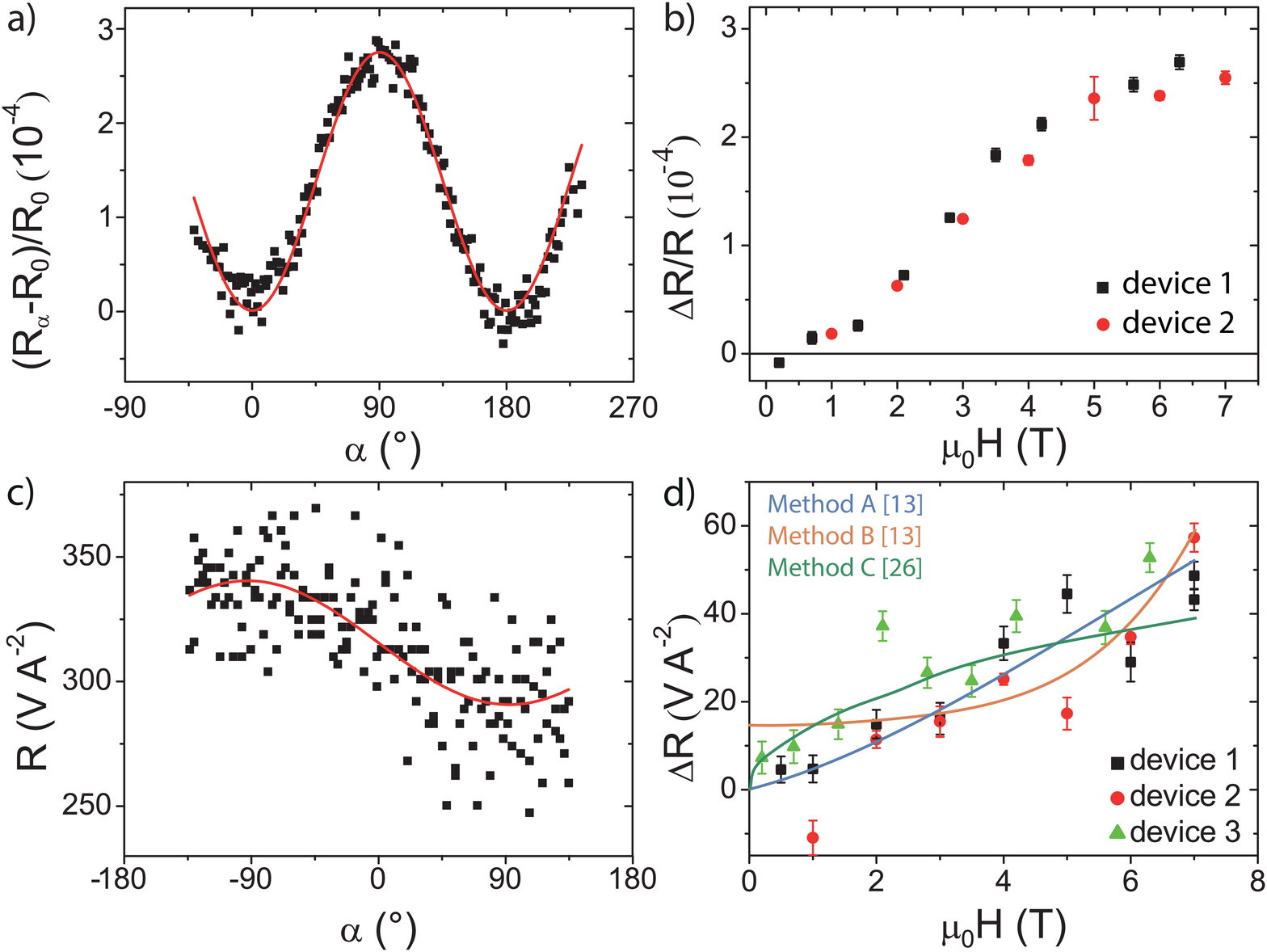}}
\end{figure*}

Easy-axis antiferromagnets have magnon modes that are typically in the THz regime. Easy-plane AFMs, however, have a more complex magnon dispersion, which are extended to lower energies. When just considering the exchange and anisotropy, a gap appears between the two modes \cite{Holanda2017a}. 
Although NiO has a simple rocksalt structure, magnetic dipole-dipole interactions and cubic anisotropy can give rise to multiple low energetic precession modes which dispersions depent on the magnetic field \cite{Milano2010}. At low temperatures (a few Kelvin), the gap from Zeeman splitting is of the order of the thermal energy and therefore could induce a non-zero spin Seebeck coefficient.

%
%
%
A magnetic field influences the spin current via the dispersion consisting of multiple modes with different magnon spin polarization. Ref. \cite{Rezende2019} treats the dispersion of the $\alpha$ and a $\beta$ modes having an offset at the zero q point even without an applied magnetic field. The offset is said to arise from the hard-axis anisotropy and is further influenced by the Zeeman interaction when applying a magnetic field. The dispersion is then given by
%

\begin{eqnarray}\nonumber
&&\omega_{\alpha,\beta}^2 / \gamma_k^2 =(H_{e}+H_{ep} +  \frac{H_{ha}}{2})^2 - \frac{1}{4} H_{ha}^2 + H^2 - H_{e}^2 \gamma_k^2 
\\
&&  \pm \left(H_{e}^2 H_{ha}^2 \gamma_k^2 + 4 H^2 ((H_{e} + H_{ep} + \frac{H_{ha}}{2})^2 - H_{e}^2 \gamma_k^2) \right)^\frac{1}{2}
\label{eq:dispersion_Rezende}
\end{eqnarray}

where $\gamma_k=\cos(\frac{1}{2}k a_l)$, H$_e$=968.4$\,$T, H$_{ha}$=635$\,$mT and H$_{ep}$=11$\,$mT are the structure factor, exchange interaction and the hard- and easy-axis anisotropy, respectively \cite{Hutchings1972, Kampfrath2011, Grimsditch1998, Rezende2019}. 
The offset causes an unequal population of these modes, especially at temperatures lower than the offset temperature of the $\alpha$ mode. 
Since their magnon spin directions are considered opposite \cite{Rezende2016}, this results in a net magnon spin required for a non-zero SSE. Fig. \ref{fig:SSE}(b) shows that the offset in the dispersions as a function of magnetic field further increases, enlarging the net magnon spin. 


However, when additionally considering the symmetry breaking magnetic dipole-dipole interactions and cubic anisotropy as done by Milano et al. \cite{Milano2010}, there appear multiple low energy modes which are shown in Fig. \ref{fig:SSE}(c) for different domains and the respective magnetic moment directions as a function of the magnetic field strength. Energetically higher modes are not considered since the measurements are performed at 5$\,$K. Under the influence of a magnetic field within a $<$111$>$ easy plane (along [110]), the [2$\overline{\rm 1}$1] and to a lesser extend the  [11$\overline{\rm 2}$] magnon modes split, the [1$\overline{\rm 1}\overline{\rm 2}$] magnon modes remain degenerate, while the [211] modes are soft and become unstable from 0.55$\,$T. A magnetic field thus causes an inbalance in the occupation of these modes which have different magnon spin direction. This leads to an inbalence in the magnon populations and opens up the opportunity to investigate these magnons with the SSE. 




When there is a net magnon spin population in a magnetic material, a temperature gradient can drive a magnon spin current $J_s=-(\sigma_m \mathbf{\nabla}\mu_m+S\mathbf{\nabla}T)$ via the SSE. The spin Seebeck coefficient $S$ has a field and a temperature dependence
.
%
%
%
The flow of magnons creates a negative magnon chemical potential $\mu_m$ near the injector. Boundary conditions at interfaces  \cite{Shan2016} and, possibly, domain walls and defects, lead to magnon accumulation and reflection resulting in a positive sign of $\mu_m$ at a distance from the injector. Shown in Fig. \ref{fig:SSE}(a) is the distribution of $\mu_m$ as a result of such reflections at the interface of a thin film of YIG on GGG.
A spin current enters a Pt strip via a finite spin mixing conductance where it is converted to a charge current by the inverse spin Hall effect.

A SSE generated spin current has been observed from NiO when grown on a ferromagnet to force the preference of one type of the many possible domains by the exchange interaction which is stronger with an uncompensated ferromagnetic $<$111$>$ layer. These domains are said to have an anisotropy induced splitting without magnetic field, having ony one type of domain would result in a non-zero SSE \cite{Rezende2017}. Without the seed layer of Py below the grown NiO, no SSE was observed as the spin currents originating from different domains would cancel \cite{Holanda2017a}. Signals in such systems show hysteresis and decrease in size when decreasing the temperature, vanishing below ~100$\,$K \cite{Ribeiro2018}.


%

%
%
%
%
%
The spin-Hall magnetoresistance (SMR) is the first harmonic response and can be obtained simultaneously with the second harmonic SSE with a lock in technique \cite{Vlietstra2014}. The SMR of the Pt injector strip is sensitive to the magnetic moments underneath it, even to the N\'eel vector in antiferromagnets \cite{Hoogeboom2017}. The SHE deflects electrons in a direction depending on their eletron spin, resulting in the accumulation of electron spins at the interface with NiO. The direction of these electron spins is affected by the interaction with the magnetic moments in the NiO via the spin transfer torque. This exchange interaction is maximal when the directions of the magnetic moments and the accumulated electron spin are perpendicular. The electron spin is reflected back into the Pt and subsequently deflected by the inverse spin Hall effect determined by the electron spin. Absoprtion of spin by the magnet thereby affects the path travelled by the electrons and influences the longitudinal resistivity $\rho_L$ of the Pt layer 
%
by \cite{Bauer2013, Hoogeboom2017}
\begin{equation}
\rho_L=\rho+\Delta\rho_{0} + \Delta\rho_{1}<1-n_x^2>
\label{rhol}
\end{equation}
where $<n_x>$  is the average of the N\'eel vector along $\hat{x}$ just below the Pt injector. This technique thus indicates the influence of the magnetic field on the magnetic moments i.e. gives information about the magnetic order and domain wall growth. \\

\begin{figure*}
\floatbox[{\capbeside\thisfloatsetup{capbesideposition={right,top},capbesidewidth=6.0cm}}]{figure}[\FBwidth]
{\caption{The non-locally obtained SSE signals. (a) The angle dependent resistance at 5$\,$K and 6.3$\,$T. 
The background resistance is small in comparison to the local background resistance. (b) The second harmonic has a negative sign at distances smaller than 500$\,$nm, similar in thin ferromagnetic films, but after the sign change the signal increases monotoneously with distance. The size of the error bars increase with distance and have a large variation between devices. (c) The signal size increases in a comparable fashion with increasing magnetic field strengt as the local signal. (d) The SSE does not increase quadratically as a function of the current. Instead, the signal shows some current threshold behavior untill 25$\,\mu$A and in increases with field untill it saturates around 75$\,\mu$A.}
	\label{fig:2nd_harmonic}}
{\includegraphics[width=9.4cm]{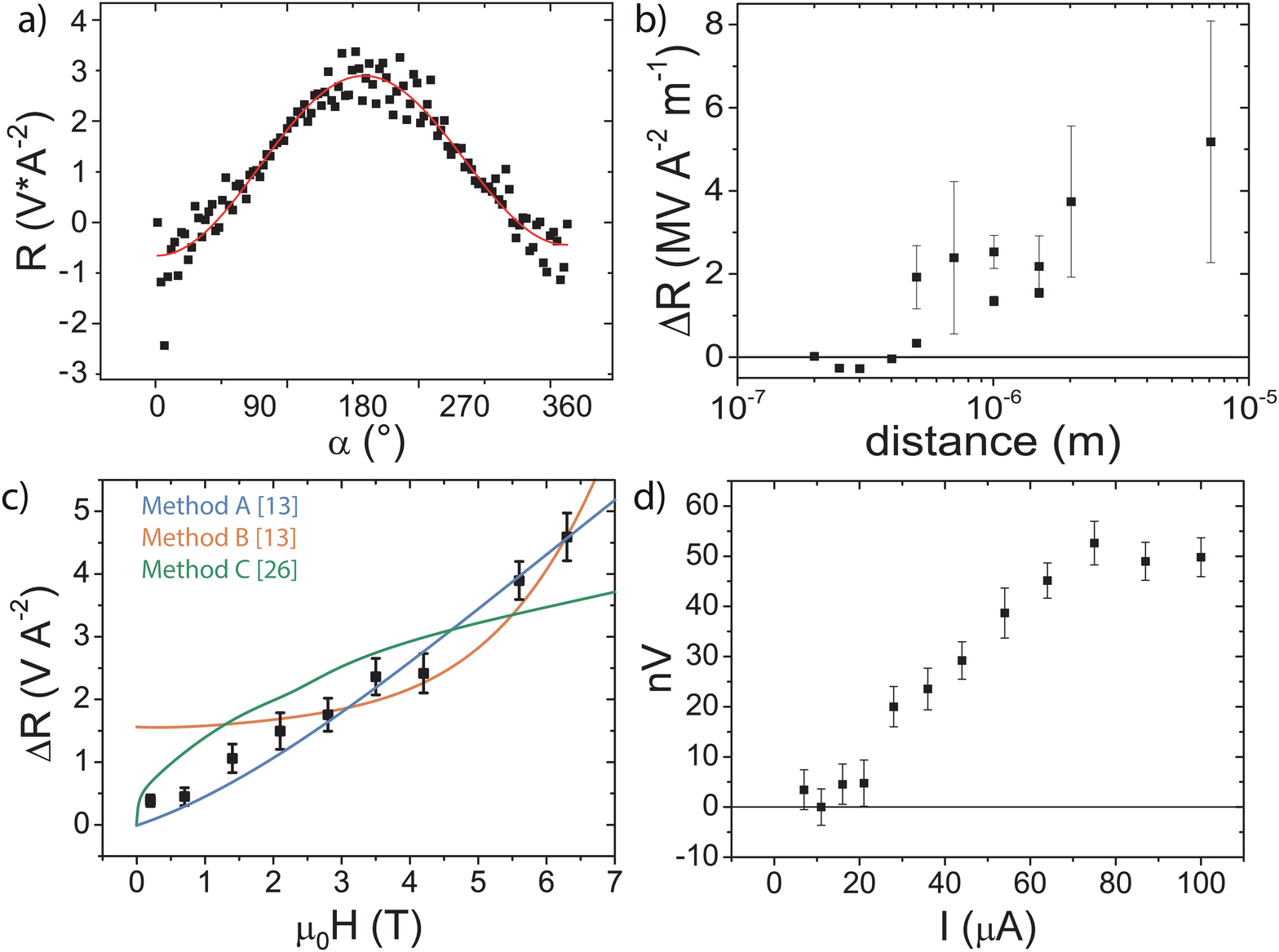}}
\end{figure*}

%

NiO is a cubic material with antiferromagnetic interaction due to superexchange between two Ni atoms via an oxygen ion. Together with magnetic dipole-dipole interactions and a small cubic anisotropy this results in the spins to align in ferromagnetic \{111\} planes which are intercoupled antiferromagnetically \cite{Milano2004}. The magnetic moments themself align along [11$\overline{\rm 2}$], slightly diverting from these \{111\} planes. Due to magnetostriction the crystal is rhombohedrically distorted along the $<$111$>$ directions and magnetic twin (T)-domains are formed \cite{Roth1960}. Within a T-domain, the three easy axes give rise to corresponding spin rotation (S)-domains. By introducing a magnetic field, the degeneracy of energetically equivalent domains is lifted resulting in a redistribution of these domains by movement of the domain walls. The direct influence on the spin rotation causes movement of the S-domains. Domain walls can influence the rotation of the the N\'eel vector and thereby both the SMR as the SSE.


The bulk NiO sample is commercially obtained and polished along a $<$111$>$ plane as described in Ref. \cite{Hoogeboom2017}. Thereafter, the devices were fabricated using electron beam lithography. No etching was performed before the sputtering of the 5$\,$nm thick, 20$\,\mu$m long and 100$\,$nm wide Pt strips. Three devices are fabricated with distances between the Pt strips of 250$\,$nm to 7$\,\mu$m. 



The SMR measurements show a $\sin^2{\alpha}$ angle dependence, see Fig. \ref{fig:locally}a). The changes in the magnetic moments are such that they tend to align perpendicular to the magnetic field direction, as to maximize the negative Zeeman energy. The signals are therefore 90$\,\degree$ angular shifted as compared to Pt$|$FM systems, confirming antiferromagnetic order in the material \cite{Hoogeboom2017}. 
Similar to our earlier work on this sample \cite{Hoogeboom2017}, the amplitude of such rotation measurements initially increase quadratically as a function of the field strength and a saturation sets in which is established around 6$\,$T, see Fig. \ref{fig:locally}b). This field dependence is believed to be originating from both anisotropy and domain wall movement. As domains with the magnetic easy plane in the $<$111$>$ surface plane become more favourite over other domains, these domains grow in size \cite{Hoogeboom2017, Fischer2018}. The domain size at small field must be smaller than the Pt strips size to follow the same field dependence as a Hall bar device. This agrees well with the observed domain size $<$1$\, \mu$m \cite{sanger2006}. A domain wall can affect magnon spin currents and therefore the distribution of magnon chemical potential as well. At saturated field strengths, the magnetic moments are coherently rotated by the magnetic field and the crystal is more or less in a single magnetic domain.


The locally measured current induced SSE shows an angular dependence and signal size which is similar to that of local measurements of Pt on FMs, see Fig. \ref{fig:locally}c). The noise of 40$\,V A^{-2}$ is relatively large in comparison to Pt$|$FM systems and might be originating from domain walls that move due to the changing magnetic field direction. There is a background signal of which the origin can be other heat related effects such as the spin-Nernst or the Righi-Leduc effect. The size of the signal amplitude as a function of the magnetic field strength shows on average a monotonous increase for all three devices as shown in Fig. \ref{fig:locally}d) apart from the substantional variation between measurements.

The SSE amplitude as a function of the magnetic field strength was fitted using the contribution of the different modes by method A and B described in Ref. \cite{Rezende2019} and method C from Ref. \cite{Rezende2018}. The magnon dispersion from Eq. \ref{eq:dispersion_Rezende} has been used in method B and the zero q-point value has been altered by the data of Ref. \cite{Milano2010} presented in Fig. \ref{fig:SSE}c) for method A and C, assuming that only small q-values play a role at the low temperature of 5$\,$K. It is assumed that the magnons are close to the thermodysanic equilibrium allowing the use the Bose-Einstein distribution for both magnon branches at a temperature of the phonon bath \cite{Troncoso2020,Rezende2019}. For method A and B, the spin Seebeck coefficient of the splitted magnon modes is given by

\begin{equation}
S_S^z=S_0 \int dk k^2 \left[\frac{e^{\frac{\hbar \omega_{\beta k}}{k_B T}}\omega_{\beta k}\nu_{\beta k y}^2}{\eta_{\beta k} (e^{\frac{\hbar \omega_{\beta k}}{k_B T}}-1)^2}-\frac{e^{\frac{\hbar \omega_{\alpha k}}{k_B T}}\omega_{\alpha k}\nu_{\alpha k y}^2}{\eta_{\alpha k} (e^{\frac{\hbar \omega_{\alpha k}}{k_B T}}-1)^2}\right]
\label{eq:SSE}
\end{equation}

where S$_0=\frac{\hbar }{6 \pi^2 k_B T^2}$, $\omega_{\mu k}$ is the field dependent frequency and $\eta_{\mu k}$ is the magnon relaxation rate of magnon mode $\mu$ as function of wave vector $k$. Further we need the magnon group velocity $\nu_{\mu k} = \delta\omega_{\mu k} / \delta k $ and the relaxation rate $\eta_q = 1 +(1240q + 5860q^3)(T/300)^3$ \cite{Rezende2018}. The values are numerically solved at a large set of field strengths which interpolating function can be fitted to the SSE amplitude data as a function of the field strength. Using the method discribed in \cite{Rezende2019} this amounts to $S_S^z = 7.8
\, 10^{-11} \,$ erg cm$^{-1}$ K$^{-1}$ for the dispersion in Eq. \ref{eq:dispersion_Rezende} (method A) and $S_S^z = 2.4
\, 10^{-11} \,$ erg cm$^{-1}$ K$^{-1}$ using with adjusted dispersion (method B) at 7$\,$T.

Both the near linear increase and the relatively small offset in the data are not represented by method B instead showing a large offset and an approximately quadratic increase with field strength. Possibly these modes disappear due to spin reorientations when applying a field. 
By following the model described in \cite{Rezende2019} and using Eq. \ref{eq:dispersion_Rezende}, the field dependence shows saturating behavior and a less significant offset. When using the dispersion including the effects of magnetic dipole-dipole interactions and cubic anisotropy, the SSE field dependence resembles the data and no offset is present. At 20$\,$K the signal has reduced to (3.2 $\,\pm\,$0.3) 10$^{-3}$ V$\,$A$^{-2}$ at 1.5$\,$T, in agreement with this explanation. 

The temperature gradient is calculated using its relation with the signal size given by \cite{Holanda2017}
\begin{equation}
V_{SSE}=R_N l \lambda_N \frac{2 e}{\hbar}\theta_{SH} \tanh{(\frac{t_N}{2 \lambda_N})} S^z_S \nabla_z T
\label{VSSE}
\end{equation}
with R$_N$=5.17$\,$k$\Omega$, l=20$\,\mu$m,  t$_N$=8$\,$nm, $\lambda_N$=1.1$\,$nm and $\theta_{SH}=0.08$ are the Pt bar resistance, length, thickness, spin diffusion length and spin Hall angle \cite{Vlietstra2013}. Further, it is assumed that the NiO thickness $\gg$ the relaxation length.  The temperature gradient along $\hat{z}$ near the injector with a current of 100$\,\mu$A is calculated to be $2.50 \, 10^6\,$ K m$^{-1}$ for method C, method A gives $7.51 \, 10^5\,$ K cm$^{-1}$ and method B results in $2.30 \, 10^3\,$ K cm$^{-1}$. This is lower compared to the calculated average temperature gradient resulting from a similar geometry on YIG at 300$\,$K of $1.6 \, 10^8\,$ K cm$^{-1}$ \cite{Shan2016} which indicates an overestimation of the calculated $S_S^z$ value. On the other hand, the thermal conductivity can be different in NiO at 5$\,$K as compared to YIG at 300$\,$K
. 
%

Fig. \ref{fig:2nd_harmonic}a) shows the angular dependence of a SSE signal obtained non-locally at the detector and is similar to that of devices on thick YIG with a strip distance in the same range \cite{Shan2016}. The distance dependence of the signal, shown in Fig. \ref{fig:2nd_harmonic}b), shows a sign change around $d \approx $ 500$\,$nm indicating that $\mu_m$ turns positive. Moreover, after the sign change, the signal seems to increase with increasing distance. In an yttrium iron garnet thin film on a gadolinium gallium garnet substrate, such sign changes are subscribed to boundary conditions at the interface of the thin film and the paramagnetic substrate. The interface will conduct little magnons while the heat is transported into GGG. Therefore, the magnons accumulate and are reflected causing $\mu_m$ to turn positive at a certain distance from the injector \cite{Shen2019}. With further increasing distances the signal in FMs drops according to a diffusion-relaxation model \cite{Cornelissen2015}. However, single domain bulk FMs lack these boundaries and no positive $\mu_m$ is observed \cite{Shen2019}. Recently, it is shown that the rotation of the pseudospin by a magnetic field could result in a sign change in case of precenence of a Dzyaloshinskii-Moriya interaction \cite{Wimmer2020} which is not expected in NiO.
It is not fully understood why this bulk NiO shows both a sign change as the following increase in signal strength with distance. 
The difference between FMs and easy-axis AFMs, however, is that in AFMs there are domains present at the relevant field strengths. 
%
%
The partial reflection and/or absorption of magnons at domain walls could give a similar upturn of  $\mu_m$ as the boundary conditions at the FM$|$PM interface. 
While $S_S$ is altered by the magetic field stregnth, this could affect $\mu_m$ such that the region where the sign change occurs is stretched. The possibility exists that the maximum reached after the sign change is then shifted towards further distances such that the signal is increasing with distance in the investigated lengthscale. Further, this hypothesis is able to explain the large fluctuations between data points which increases with increasing distance being due to movement of domain walls creating local variation in $\mu_m$. 

Fig. \ref{fig:2nd_harmonic}c) shows the magnetic field strength dependence, fitted with the same models are used as for the local data, assuming that the S$_S$ is dominant for the change in signal strength. Also for the non-local signals the approach described in Ref. \cite{Rezende2019} using the dispersion from Ref. \cite{Milano2010} best resembles the near-linear field dependence without offset. The SSE signal is driven by Joule heating and therefore expected to have a quadratic dependence on the current send through the injector. However, no clear quadratic current dependence is observed in Fig. \ref{fig:2nd_harmonic} and instead a saturation sets in around 75$\,\mu$A. A temperature increase by the current can slightly lower the signal, being strongly temperature dependent. 
%
Both the field dependence and the absence of an electrically injected signal resembles the results reported for Cr$_2$O$_3$ by Yuan et al. \cite{Yuan2018}. However, they claim this spin transport is a result of spin superfluidity.

The SSE originates from influence of a magnetic field on the population of the magnon modes but these models might be influenced by the movement of domain walls. The lacking offset in method B could be explained by the multi domain nature as the domains have opposite magnon spin polarization resulting ina smaller net SSE. 
Moreover, domain walls can interact with the spin current itself leading to domain wall movement \cite{Wieser2011} and spin current reflection upon domain walls \cite{Tveten2013}. In thin films with multiple domain walls, the reflection damps the non-local signals, but domain optimization by tuning the growth direction or by magnetic training still leads to micrometer spin transport \cite{Ross2020}. 
To conclude, we observed a spin Seebeck effect generated spin current in the bulk easy-plane antiferromagnet NiO. 
This is achieved as a result of an applied magnetic field without the need of a exchange interactions to align the magnetic moments.
The field dependence of the SSE amplitude at 5$\,$K is modelled by the energy splitting of magnon modes, creating an imbalance in the magnon spin population.
The cubic anisotropy and magnetic dipole-dipole interactions have to be taken into account in order to recreate the near linear SSE dependence on the field and the small offset. 
Furthermore, the SSE signal exhibits both a sign change and then an increase with increasing the injector distance that would not be possible without the introduction of additional boundary conditions in the bulk NiO, a role that may have been fulfilled by domain walls.



We acknowledge J. G. Holstein, H. Adema, T. J. Schouten and H. H. de Vries for their technical assistance. Further, we thank Rembert Duine and Camillo Ulloa for fruitfull discussions. Support by the research program Magnon Spintronics (MSP) No. 159 financed by the Nederlandse Organisatie voor Wetenschappelijk Onderzoek (NWO) and the Spinoza prize awarded to Professor B. J. van Wees by the NWO is gratefully acknowledged.

\bibliography{references2}

\end{document}